\documentclass[aip, preprint, twocolumn, amsmath, amssymb, reprint, superscriptaddress,floatfix]{revtex4-1}

\newcommand{\be}{\begin{equation}}
	\newcommand{\ee}{\end{equation}}
\newcommand{\bea}{\begin{eqnarray}}
	\newcommand{\eea}{\end{eqnarray}}
\newcommand{\Fig}[1]{Fig.\,\ref{#1}}
\newcommand{\Eq}[1]{Eq.\,(\ref{#1})}
\newcommand{\etal}{{\it et al. }}

\usepackage{graphicx,verbatim}
\usepackage{dcolumn}
\usepackage{bm}
\usepackage{bbm}
\usepackage{sidecap}
\usepackage{braket}
\usepackage{color}
\usepackage{amssymb}
\usepackage{MnSymbol}
\usepackage{csquotes}
\usepackage{pifont}

\newcommand*{\rom}[1]{\expandafter\@slowromancap\romannumeral #1@}

\begin{document}
\title{
Tree tensor network state approach for solving hierarchical equations of motion
}

\author{Yaling Ke}\email{yaling.ke@physik.uni-freiburg.de}
\affiliation{Institute of Physics, University of Freiburg, Hermann-Herder-Strasse 3, 79104 Freiburg, Germany}

\begin{abstract}
The hierarchical equations of motion (HEOM) method is a numerically exact open quantum system dynamics approach. The method is rooted in an exponential expansion of the bath correlation function, which in essence strategically reshapes a continuous environment into a set of effective bath modes that allow for more efficient cutoff at finite temperatures. Based on this understanding, one can map the HEOM method into a Schrödinger-like equation, with a non-Hermitian super Hamiltonian for an extended wave function being the tensor product of the central system wave function and the Fock state of these effective bath modes. In this work, we explore the possibility of representing the extended wave function as a tree tensor network state (TTNS), the super Hamiltonian as a tree tensor network operator of the same structure as the TTNS, as well as the application of a time propagation algorithm using the time-dependent variational principle. Our benchmark calculations based on the spin-boson model with a slow-relaxing bath show that the proposed HEOM+TTNS approach yields consistent results with that of the conventional HEOM method, while the computation is considerably sped up. Besides, the simulation with a genuine TTNS is four times faster than a one-dimensional matrix product state decomposition scheme. 
 \end{abstract}	
\maketitle
The efficient simulation of open quantum system dynamics lies at the heart of a great variety of chemical and physical applications,\cite{Breuer__2002_p,May__2008_p,Weimer_Rev.Mod.Phys._2021_p015008} ranging from efficient energy and charge transfer in natural and artificial photosynthetic systems to quantum transport in nanoscale electronic devices.\cite{Ishizaki_ProceedingsoftheNationalAcademyofSciences_2009_p17255,Bredas_Nat.Mater._2017_p35,Thoss_J.Chem.Phys._2018_p30901,Cuevas__2010_p} Over the past few decades, the hierarchical equations of motion (HEOM) method has grown into a mainstream non-perturbative and non-Markovian open quantum system dynamics approach.\cite{Tanimura_J.Phys.Soc.Jpn._1989_p101,Yan_Chem.Phys.Lett._2004_p216,Xu_Phys.Rev.E_2007_p31107,Jin_J.Chem.Phys._2008_p234703,Yan_J.Chem.Phys._2014_p54105,Schinabeck_Phys.Rev.B_2018_p235429,Hsieh_J.Chem.Phys._2018_p14103,Ye_WIREsComputMolSci_2016_p608,Tanimura_J.Chem.Phys._2020_p20901}
The core idea of the method is based on an exponential series expansion of the bath correlation function as well as successively taking time-derivative of the reduced system density operator in the influence functional formalism, so as to construct a group of auxiliary density operators (ADOs) that obey a hierarchical set of differential equations. 

While the HEOM method has gained tremendous success aided by abundant optimization schemes and advances in computational architecture,\cite{Ishizaki_J.Phys.Soc.Jpn._2005_p3131,Shi_J.Chem.Phys._2009_p84105,Hu_J.Chem.Phys._2010_p101106,Struempfer_J.Chem.TheoryComput._2012_p2808,Tsuchimoto_J.Chem.TheoryComput._2015_p3859,Kramer_J.Comput.Chem._2018_p1779,Chen_J.Chem.Phys._2022_p221102,Xu_Phys.Rev.Lett._2022_p230601,Ikeda_J.Chem.Phys._2022_p104104} its applications hit a hard wall in the cases where the central system size is large,  many exponential terms (or effective bath modes) are required to reproduce the original bath correlation function, and a deep hierarchical depth is necessitated to account for the strong system-bath coupling accurately. This is because the computational cost grows exponentially with the increase of these factors.
Recently, these limitations have been effectively broached by employing the matrix product states (MPS),\cite{Schollwoeck_Ann.Phys.NY_2011_p96--192} also known as tensor trains (TT) decomposition\cite{Oseledets_SIAMJ.Sci.Comput._2011_p2295--2317} of these ADOs.\cite{Shi_J.Chem.Phys._2018_p174102,Borrelli_J.Chem.Phys._2019_p234102,Yan_J.Chem.Phys._2020_p204109,Ke_J.Chem.Phys._2022_p194102} MPS/TT provides a compact and ideal way to encode one-dimensional (1D) short-range quantum many-body system correlation,\cite{Vidal_Physicalreviewletters_2004_p40502} by representing the multi-dimensional wave function as the product of a chain of low-rank tensors. When the entanglements are well confined within the near neighboring tensors over a long time, the computational cost of the HEOM+MPS/TT method scales linearly with the system size, the number of effective bath modes, and the hierarchical depth. However, it is not often the case for the HEOM+MPS/TT method, because in addition to the strong correlation among system degrees of freedom (DoF),  the system DoF may also be strongly coupled to every effective bath mode. As a consequence, the correlation would quickly spread over a long distance along the MPS chain, and very large bond dimensions are needed to obtain accurate results. Therefore, it is important to go beyond the MPS/TT ansatz and explore the combination of the HEOM method with other higher-dimensional tensor network state structures that can more efficiently encode the inherent entanglement between the system DoFs and effective bath modes.

In this work, we will demonstrate the applicability of tree tensor network state (TTNS) which is a generalization of the MPS/TT into a tree-shaped network of tensors,\cite{Shi_Phys.Rev.A_2006_p22320,Tagliacozzo_Phys.Rev.B_2009_p235127,Murg_Phys.Rev.B_2010_p205105,Li_Phys.Rev.B_2012_p195137,Changlani_Phys.Rev.B_2013_p85107,Nakatani_J.Chem.Phys._2013_p134113,Murg_J.Chem.TheoryComput._2015_p1027,Gunst_J.Chem.TheoryComput._2018_p2026,Schroeder_Nat.Commun._2019_p1,Larsson_J.Chem.Phys._2019_p204102,Ferrari_Phys.Rev.B_2022_p214201,Seitz_arXivpreprintarXiv2206.01000_2022_p,Milsted_arXivpreprintarXiv1905.01331_2019_p}  and a time propagation algorithm based on the time-dependent variational principle (TDVP)\cite{Bauernfeind_SciPostPhysics_2020_p024,Kloss_SciPostPhysics_2020_p70,Ceruti_SIAMJ.Numer.Anal._2021_p289} for solving the HEOM method.

In general, an open quantum system describes a central system of interest coupled to a macroscopic environment, and the Hamiltonian reads
\begin{equation}
H = H_S + H_B + H_{SB},
\end{equation}
where $H_S$, $H_B$, and $H_{SB}$ correspond to the system, bath Hamiltonian, and their interaction, respectively. As a demonstration of the concept, we start by considering a simple but paradigmatic open quantum system, the spin-boson model,  where the system Hamiltonian is given by
\begin{equation}
\label{ori-model}
H_S =  \epsilon \sigma_z + \Delta \sigma_x. 
\end{equation}
Here, the energy bias between two spin states is given by $2\epsilon$
and $\Delta$ denotes the coupling between two states. The spin is coupled to a dissipative bosonic environment, modeled as a phonon bath comprised of an infinite number of harmonic oscillators,
\begin{equation}
\label{Hb}
H_B =   \sum_{j} \left(\frac{p_{j}^2}{2}+\frac{1}{2}\omega_{j}^2q_{ j}^2\right),
\end{equation}
where $q_{j}$ and $p_{j}$ are the mass-weighted position and conjugated momentum operator of the $j$th harmonic oscillator, and $\omega_{j}$ is the corresponding frequency. 
The system-bath coupling Hamiltonian $H_{SB}$ takes a linear form with respect to the bath coordinate $q_{j}$, 
\begin{equation}
\label{Hsb}
H_{SB} = \sigma_z\sum_{j}  c_{j}q_{j},
\end{equation}
and the system-bath coupling operator in the system subspace is given by $\sigma_z$, the coupling strength specified by $c_j$.

The composite system-bath dynamics are described by the density operator $\rho(t)$. Note that, we assume the system and bath are factorized at the initial moment and the bath is in its own thermal equilibrium state at inverse temperature $\beta$. Then, the initial density operator $\rho(0)$ is given by 
\begin{equation}
\rho(0) = \rho_S(0)\otimes\frac{e^{-\beta H_{B}}}{\mathrm{Tr}_B\{e^{-\beta H_B}\}}.
\end{equation}
In this case, in the reduced system dynamics $\rho_S(t) = \mathrm{tr}_B\{\rho(t)\}$ where all environmental DoFs are traced out, the influence of the environment on the system is characterized statistically in a time correlation function
\begin{equation}
\label{correlationfunction_ori}
C(t) 
= \frac{1}{\pi}\int_{-\infty}^{\infty} \frac{e^{-i\omega t}}{1-e^{-\beta \omega}} J(\omega) \mathrm{d}\omega.
\end{equation}
The spectral density function $J(\omega)$ encodes the coupling-weighted density of states of the bath in frequency space, 
 defined by 
\begin{equation}
J(\omega) = \frac{\pi}{2}\sum_j \frac{c_{j}^2}{\omega_{j}}\delta(\omega-\omega_{j}).
\end{equation}
For some specific forms of $J(\omega)$, such as Drude-Lorentz spectral density function,
 \begin{equation}
J(\omega) = 2\lambda\frac{\omega \Omega}{\omega^2+\Omega^2},
\end{equation}
where $\Omega$ is the bath characteristic frequency and $\lambda$ quantifies the system-bath coupling strength, the bath correlation function in \Eq{correlationfunction_ori} can be expanded analytically into an exponential summation,
\begin{equation}
\label{correlationfunction_exp}
    C(t) =  \sum_{p=0}^{\infty} \lambda\eta_{p} e^{-i\gamma_{ p}t}. 
\end{equation}
The explicit expressions of $\eta_{p}$ and $\gamma_{p}$ depend on the sum-over-pole decomposition scheme of the Bose distribution function $f(\omega) = 1/(1-e^{-\beta \omega})$. Throughout this work, we adopt the Pad\'e pole decomposition scheme \cite{Hu_J.Chem.Phys._2010_p101106,Hu_J.Chem.Phys._2011_p244106} 
and the explicit expressions for  $\eta_{p}$ and $\gamma_{p}$ can be found in the supporting information (SI). At finite temperatures, the infinite summation in \Eq{correlationfunction_exp} can be cut off at a finite $P$, which is chosen large enough to well reproduce the original bath correlation function. 
For an arbitrary form of the spectral density function, the exponential expansion analogous to \Eq{correlationfunction_exp} can be implemented numerically.\cite{Ikeda_J.Chem.Phys._2020_p204101,Chen_J.Chem.Phys._2022_p221102,Xu_Phys.Rev.Lett._2022_p230601} 

In fact, the exponential expansion in \Eq{correlationfunction_exp} lays the groundwork for the derivations of the HEOM method.  For a detailed derivation of the method, we refer readers to a review Ref. \onlinecite{Tanimura_J.Chem.Phys._2020_p20901} and the references therein. Here, we only briefly introduce the main concept of the method. The key is to introduce a group of ADOs, $\rho^{\bm{n}}(t)$, where the superscript $\bm{n}= (n_{0}, \cdots, n_{p},\cdots, n_{P})$ is an ordered set of $P+1$ non-negative integers and each $n_{p}$ is associated with a frequency component $\gamma_{p}$ in \Eq{correlationfunction_exp}. These ADOs are closed with respect to the time-derivative operation, which yields

\begin{figure*}
    \begin{minipage}[c]{0.28\textwidth} 		
	\raggedright a) 
	\includegraphics[width=\textwidth]{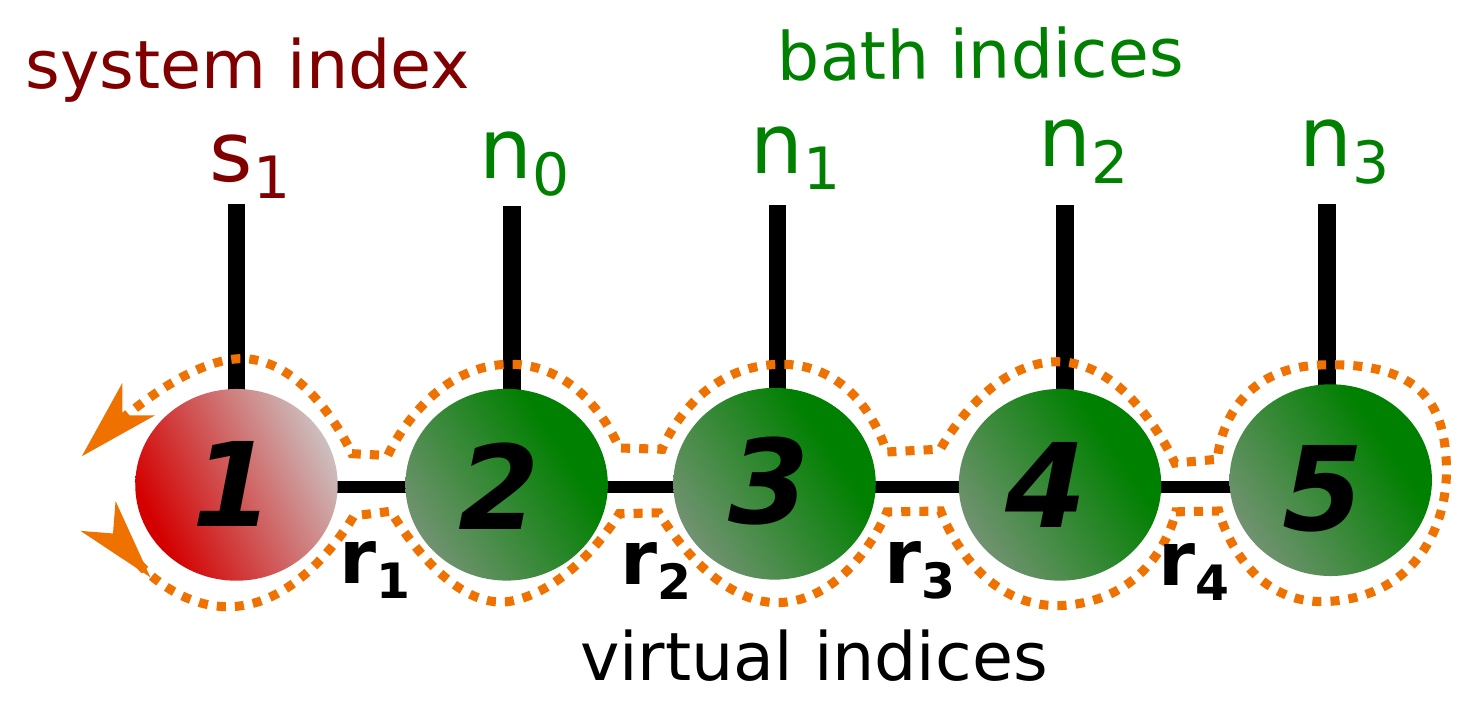}
 	\raggedright b) 
	\includegraphics[width=\textwidth]{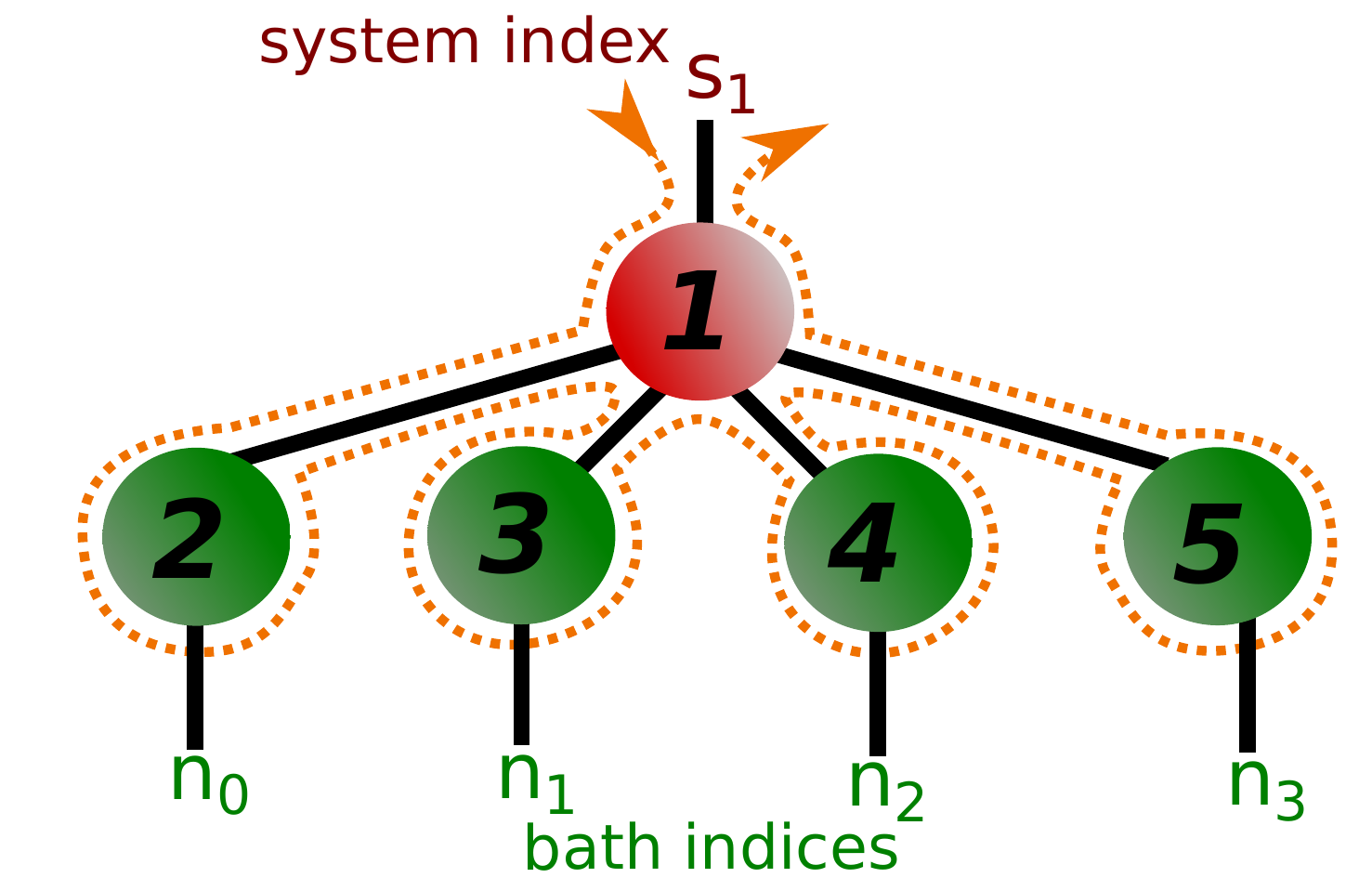}
    \end{minipage}
        \begin{minipage}[c]{0.3\textwidth} 		
	\raggedright c) 
	\includegraphics[width=\textwidth]{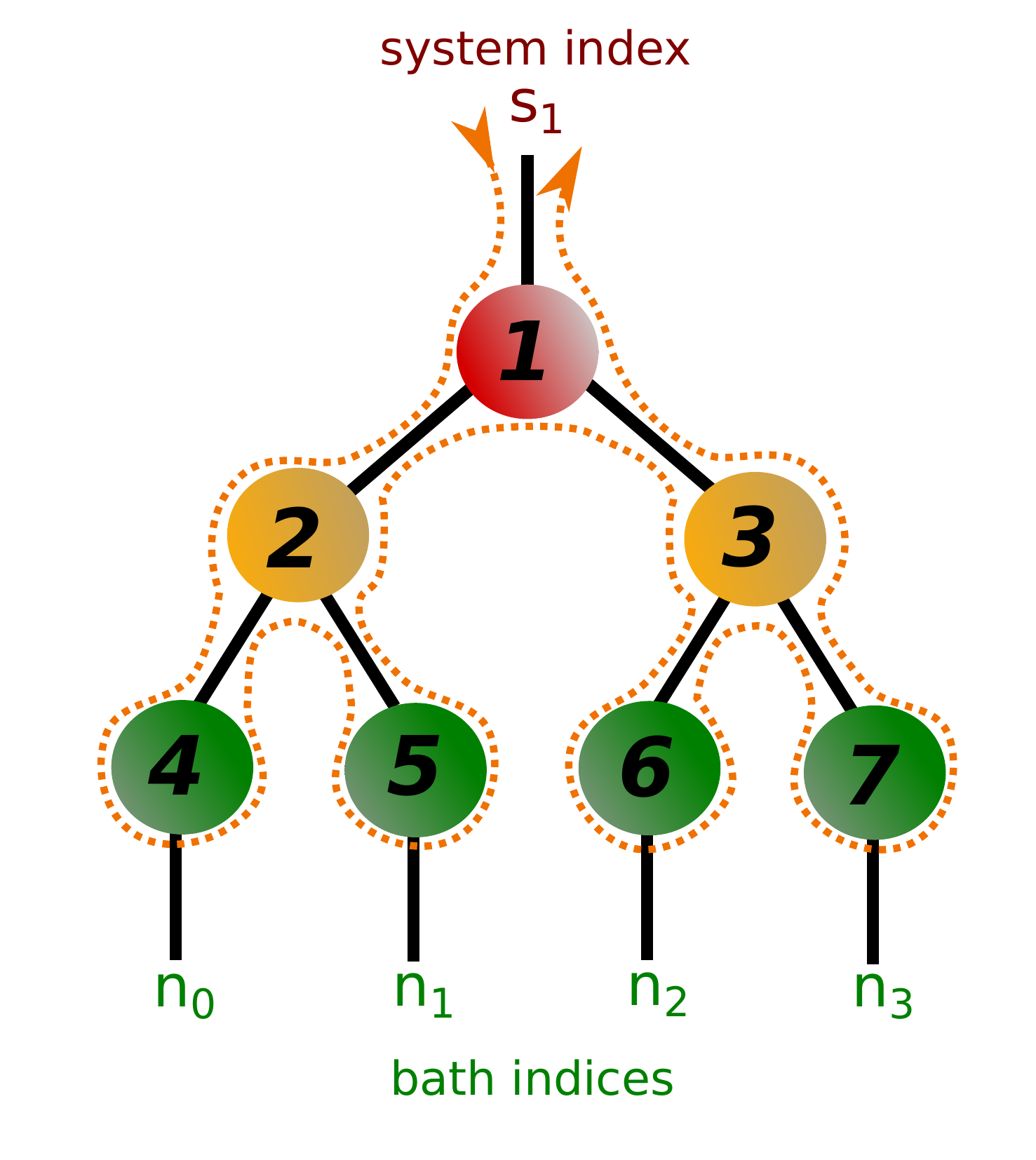}
    \end{minipage}
        \begin{minipage}[c]{0.3\textwidth} 		
	\raggedright d) 
	\includegraphics[width=\textwidth]{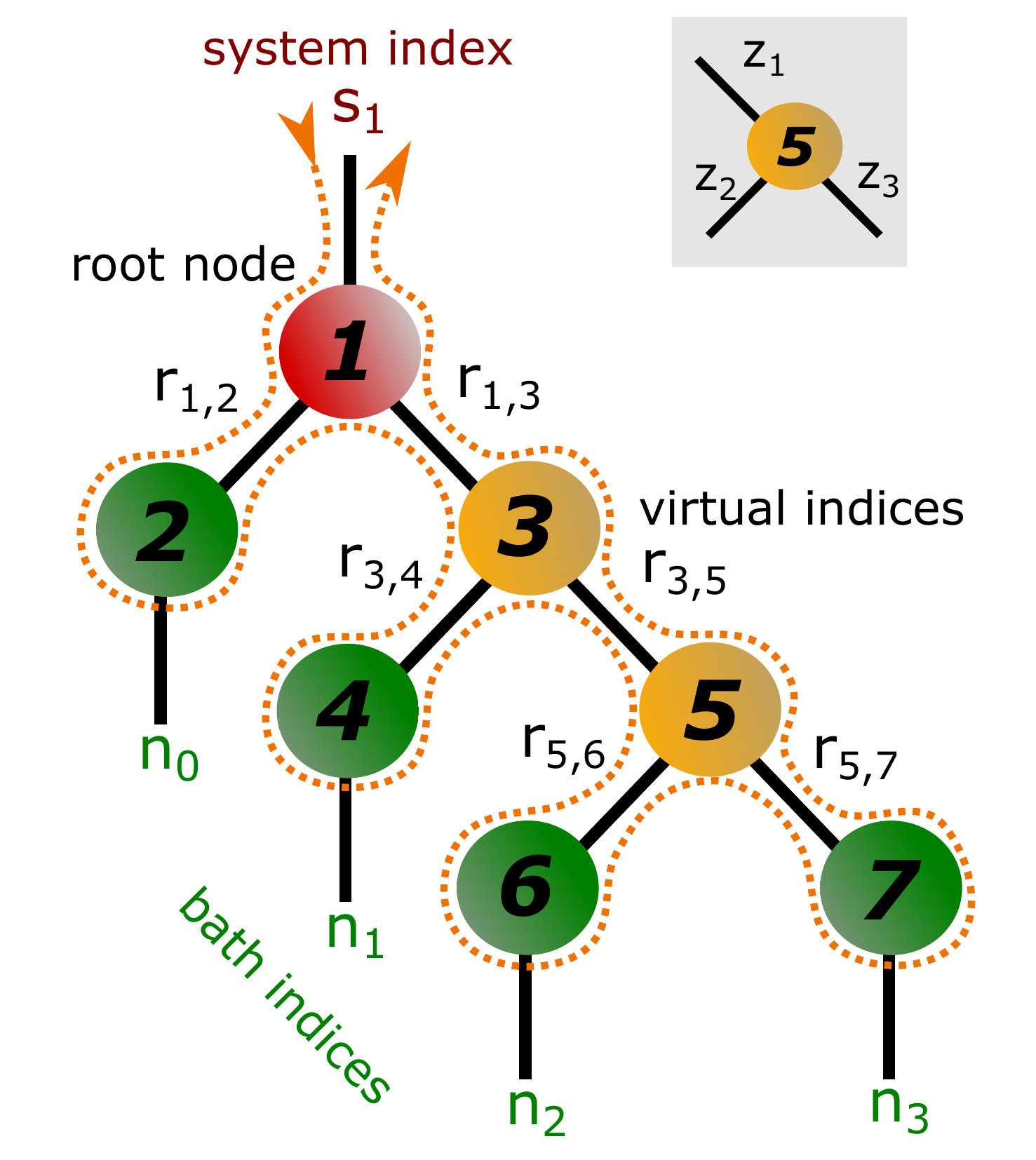}
    \end{minipage}
 \caption{Examples of the MPS/TT (a) and three TTNS decompositions (b, c, and d) in the graphical notation of the extended wave function $|\Psi(t)\rangle$ (see \Eq{ExtendedWaveFunction}) with $P=3$ for the spin-boson model. The circle nodes represent the tensors, black dangling lines denote the physical indices, and shared links correspond to virtual indices.  The orange dotted lines with arrows specify the ordered sequence sweeping through all nodes in the entire TTNS, which start and end at the root node. }
 \label{TTNS_spinboson}	
\end{figure*}
\begin{equation}
\label{heom}
\begin{split}
i\frac{d \rho^{\bm{n}}(t) }{dt} 
=&H_{S}\rho^{\bm{n}}(t)-\rho^{\bm{n}}(t)H_{S}
-i\sum_{p=0}^P n_{p}\gamma_{p}  \rho^{\bm{n}}(t) \\
 &+\sum_{p=0}^P  \sqrt{\lambda(n_{p}+1)} \left( \sigma_z \rho^{\bm{n}^+_{ p}}(t)-
   \rho^{\bm{n}_{ p}^+} (t)\sigma_z\right)\\
& +\sum_{p=0}^P \sqrt{\lambda n_{ p}}\left(\eta_{p} \sigma_z \rho^{\bm{n}_{ p}^-}(t)  -\eta_{ p}^{*} \rho^{\bm{n}^-_{p}}(t)  \sigma_z\right),
\end{split}
\end{equation}
where $\bm{n}_{p}^{\pm} = (n_{0}, \cdots, n_{p}\pm 1,\cdots, n_{P})$.  Directly propagating \Eq{heom} with a certain hierarchy truncation scheme is termed as the conventional HEOM method. The simplest truncation scheme is to set $\rho^{\bm{n}}(t) = 0$ when $\sum_{p} n_{p} > L$ and $L$ is the hierarchical truncation tier. Using this truncation scheme, the number of elements to be stored and propagated is $\frac{(P+1+L)!}{ (P+1)!L!}d^{2N_s}$, where $N_s$ is the number of system DoFs and $d$ is the size of basis set per system DoF (for the spin-boson model, we have $N_s=1$ and $d=2$). As such, the conventional HEOM method is formidably expensive in the cases of many bath poles, a deep truncation tier, and a large central system. An alternative way to reduce the computational cost and memory requirement is to employ tensor network states. To this end, it is opportune to reformulate the above hierarchical set of differential equations into a Schr\"odinger-like equation,\cite{Borrelli_J.Chem.Phys._2019_p234102,Ke_J.Chem.Phys._2022_p194102}
\begin{equation}
\label{SchroedingerEquation}
i\frac{d |\Psi(t)\rangle }{dt} =\mathcal{H} |\Psi(t)\rangle,
\end{equation}
via the introduction of an effective phonon bath and the density matrix purification scheme.\cite{Schmutz_ZeitschriftfurPhysikBCondensedMatter_1978_p97,Suzuki_Internat.J.ModernPhys.B_1991_p1821,Arimitsu_Prog.Theor.Phys._1987_p32--52,Feiguin_Phys.Rev.B_2005_p220401,Verstraete_Phys.Rev.Lett._2004_p207204,Borrelli_WIREsComputMolSci_2021_pe1539}

First, $|n_{p}\rangle$ can be interpreted as the basis in the occupation number representation of a virtual dissipative harmonic oscillator. A pair of creation and annihilation operators, $b^{\dagger}_{p}$ and $b_{p}$ are introduced, and they
act on $|n_{p}\rangle$ to yield
\begin{subequations}
\label{BathEffectiveOperators}
\begin{align}
    b^{\dagger}_{p}&|n_{p}\rangle = \sqrt{n_{p}+1}|n_{p}+1\rangle,\\
    b_{p}&|n_{p}\rangle = \sqrt{n_{ p}}|n_{p}-1\rangle.
\end{align}
\end{subequations}
All these virtual modes constitute an effective phonon bath, which encrypts the statistical information of the original bath as well as its influence on the system dynamics (see Eqs. (\ref{Hb}) and (\ref{Hsb})).
When all these effective modes are placed in their ground states, i.e., for the particular configuration of the Fock state $|\bm{n}\rangle=|0, \cdots, 0\rangle$, the reduced system density operator is reproduced, $\rho_S(t) = \rho^{(0, \cdots, 0)}(t)$. Certain bath-related properties can be extracted from the higher-order ADOs. \cite{Zhu_J.Chem.Phys._2012_p194106,Song_Phys.Rev.B_2017_p64308,Kato_J.Chem.Phys._2016_p224105,Schinabeck_Phys.Rev.B_2018_p235429} 

Second, each ADO is usually expressed as a density matrix in the Hilbert space, but it can also be recast into a rank-$N_s$ tensor,
\begin{equation}
\label{ado_tensor}
   | \rho^{\bm{n}}(t) \rrangle \equiv \sum_{s_1\cdots s_{N_s}} C^{\bm{n}}_{s_1 \cdots s_{N_s}}(t)|s_1 \cdots s_{N_s}\rangle.
\end{equation}
where $s_i$ ranges from $1$ to $d^2$. For more theoretical and technical details with regard to this space transformation, we refer readers to Refs. \onlinecite{Schmutz_ZeitschriftfurPhysikBCondensedMatter_1978_p97,Suzuki_Internat.J.ModernPhys.B_1991_p1821,Arimitsu_Prog.Theor.Phys._1987_p32--52,Feiguin_Phys.Rev.B_2005_p220401,Verstraete_Phys.Rev.Lett._2004_p207204,Borrelli_WIREsComputMolSci_2021_pe1539}.

All the ADOs combined comprise an extended wave function in the enlarged space
\begin{equation}
\label{ExtendedWaveFunction}
|\Psi(t)\rangle=\sum_{\begin{subarray}{c}  s_1 \cdots s_{N_s} \\  n_{0}\cdots n_{P} \end{subarray}}
    C^{\bm{n} }_{\bm{s}}(t)   |s_1\cdots  s_{N_s}\rangle \otimes |n_{0} \cdots n_{P}\rangle.
\end{equation}
The reduced system density operator is obtained by the partial product $
|\rho_S(t)\rrangle \equiv \langle \bm{n}=\bm{0}|\Psi(t)\rangle$. A system observable $\langle O(t)\rangle$ is obtained by 
\begin{equation}
\langle O(t)\rangle = \mathrm{tr}_S\left\{O\rho_S(t)\right\}
=\langle \bm{n}=\bm{0}| \bigotimes_{i=1}^{N_s} \langle \vec{\mathbbm{1}}_d^i| \hat{O} |\Psi(t)\rangle,
\end{equation}
where $O$ can be any operator in the system subspace, and $\hat{O}=O\otimes \mathbbm{1}_d$ with $\mathbbm{1}_d$ being an $d\times d$ unit matrix. $\vec{\mathbbm{1}}_d$ is a vector of length $d^2$, obtained by vectorizing $\mathbbm{1}_d$. For instance,  $\vec{\mathbbm{1}}_2=(1\,\, 0 \,\, 0 \,\,1)^{\text{transpose}}$.

The super Hamiltonian $\mathcal{H}$ in \Eq{SchroedingerEquation} can be written out explicitly with the operators introduced in \Eq{BathEffectiveOperators} as 
\begin{equation}
\label{SuperHamiltonian}
\begin{split}
\mathcal{H} =&\mathcal{H}_S -i\sum_{p=0}^P \gamma_{p} b^{\dagger}_{p}   b_{p}  
+\sqrt{\lambda}\hat{\sigma}_z
\sum_{p=0}^P  \left( b_{p} + \eta_{p}  b^{\dagger}_{p} \right) \\
&-\sqrt{\lambda} \tilde{\sigma}_z \sum_{p=0}^P  \left(  b_{p}   +\eta_{p}^{*}  b^{\dagger}_{p} \right), \end{split}
\end{equation}
where $\mathcal{H}_S=\epsilon \left( \hat{\sigma}_z - \tilde{\sigma}_z\right) + \Delta \left( \hat{\sigma}_x - \tilde{\sigma}_x \right) $ with 
$\hat{\sigma}_{z/x} =  \sigma_{z/x}\otimes \mathbbm{1}_2$, $\tilde{\sigma}_{z/x}= \mathbbm{1}_2 \otimes \sigma_{z/x}$, and $\mathbbm{1}_2$ being an $2\times 2$ unit matrix. The super operator $\mathcal{H}_s$ originates from the space transformation of $H_S$ in correspondence with \Eq{ado_tensor}.
It is worth noting that the super Hamiltonian $\mathcal{H}$ is non-Hermitian. As such, the norm of the extended wave function $|\Psi(t)\rangle$ is not conserved. 

The direct time integration of \Eq{SchroedingerEquation} along with the explicit definition of the extended wave function $|\Psi(t)\rangle$ and super Hamiltonian $\mathcal{H}$ given in \Eq{ExtendedWaveFunction} and \Eq{SuperHamiltonian}, respectively, is usually an intractable task as the size of the coefficient tensor $C^{\bm{n}}_{\bm{s}}(t)$ scales exponentially with the number of extended system DoFs, $N=N_s+P+1$. This huge amount of data can be compressed in the format of a tensor network state, which is a structured product of low-rank tensors.

The well-established MPS/TT format is a special and widely-used instance of tensor network states, where the tensors are arranged into a one-dimensional chain.\cite{Schollwoeck_Ann.Phys.NY_2011_p96--192,Oseledets_SIAMJ.Sci.Comput._2011_p2295--2317,Cirac_J.Phys.AMath.Theor._2009_p504004,Chan_J.Chem.Phys._2016_p14102,Jaschke_Comput.Phys.Commun._2018_p59} The high-rank coefficient tensor $C^{\bm{n}}_{\bm{s}}(t)$ in the extended wave function $|\Psi(t)\rangle$ can be approximated in the MPS/TT formalism as 
\begin{equation}
\label{MPS}
\begin{split}
C_{\bm{s}}^{\bm{n}}(t) &\approx \sum_{ r_1  \cdots r_{N-1}}    
        A^{[1]}_{r_{1},s_1} \cdots     A^{[N_s+1]}_{r_{N_s},r_{N_s+1},n_{0}}  \cdots A^{[N]}_{r_{N-1}, n_{P}} \\
        &\approx \mathrm{Contr}\left\{ A^{[1]}_{s_1} \cdots  A^{[N_s+1]}_{n_{0}}  \cdots A^{[N]}_{n_{P}} \right\}.
\end{split}
\end{equation}
Here, $A^{[i]}$ are rank-3 tensors with one physical index ($s_i$ or $n_{p}$) and two virtual indices $r_{i-1}$ and $r_i$, except for $A^{[1]}$ and $A^{[N]}$, which are rank-2 tensors and thus have only one virtual index. The virtual index $r_i$ runs from 1 to $D_i$, where the bond dimension $D_i$ is a controllable parameter, that can be systematically increased to reduce the degree of approximation in \Eq{MPS}. The maximum value among $\{D_i\}$ is designated as the maximal bond dimension $D_{\mathrm{max}}$. An example of the MPS/TT representation of $|\Psi(t)\rangle$ with $P=3$ is illustrated graphically in  \Fig{TTNS_spinboson} (a). The circle nodes correspond to the tensors. The dangling legs denote the physical indices, and the connected link between two neighboring nodes is assigned with a virtual index $r_i$. The summation over a virtual index shared between two tensors is called the contraction. Contracting all virtual indices (denoted in \Eq{MPS} as $\mathrm{Contr}$) reproduces approximately $C_{\bm{s}}^{\bm{n}}(t)$.

The TTNS is an extension of the MPS/TT beyond 1D geometry and thus can accommodate a more complicated entanglement network.\cite{Shi_Phys.Rev.A_2006_p22320,Tagliacozzo_Phys.Rev.B_2009_p235127,Murg_Phys.Rev.B_2010_p205105,Li_Phys.Rev.B_2012_p195137,Changlani_Phys.Rev.B_2013_p85107,Nakatani_J.Chem.Phys._2013_p134113,Murg_J.Chem.TheoryComput._2015_p1027,Gunst_J.Chem.TheoryComput._2018_p2026,Schroeder_Nat.Commun._2019_p1,Larsson_J.Chem.Phys._2019_p204102,Ferrari_Phys.Rev.B_2022_p214201,Seitz_arXivpreprintarXiv2206.01000_2022_p,Milsted_arXivpreprintarXiv1905.01331_2019_p} It is also called hierarchical Tucker format,\cite{Lubich_SIAMJ.MatrixAnal.Appl._2013_p470} which is the ansatz behind an important quantum dynamics approach, multi configuration time-dependent Hartree and its multi-layer variant.\cite{Beck_ZeitschriftfurPhysikDAtomsMoleculesandClusters_1997_p113,Wang_J.Chem.Phys._2003_p1289,Meyer__2009_p,Manthe_J.Chem.Phys._2015_p244109,Wang_J.Phys.Chem.A_2015_p7951,Lindoy_J.Chem.Phys._2021_p174108} A general TTNS connects all tensors into a tree structure  and the network is loop-free,\cite{Montangero_Philos.Trans.R.Soc.A_2022_p20210065} which means that there exists a unique path between any two tensors. 
A TTNS representation for $C_{\bm{s}}^{\bm{n}}(t)$ in the extended wave function $|\Psi(t)\rangle$ is directly analogous to \Eq{MPS}, 
\begin{equation}
\label{TTNS}
\begin{split}
C_{\bm{s}}^{\bm{n}}(t) \approx \mathrm{Contr}\left\{ T^{[1]}_{s_1} T^{[2]}_{n_0}T^{[3]}_X\cdots  T^{[k]} \cdots T^{[K]}_{n_P} \right\},
\end{split}
\end{equation}
except that $T^{[k]}$ is now a rank-$z$ tensor, where $z$ can be higher than 3. 

There are many different ways to connect the tensors into a tree-shaped network. For example, we show in \Fig{TTNS_spinboson} the graphical representation of several different TTNS decompositions for the extended wave function $|\Psi(t)\rangle$ with $P=3$. A circle node with $z$ legs corresponds to a rank-$z$ tensor. For convenience, the nodes are enumerated and we take the TTNS in \Fig{TTNS_spinboson} (d) as an example to introduce the notations that will be used in what follows. Similar to an MPS/TT, there are nodes assigned with one dangling leg denoting the physical index $s_1/n_{p}$, such as nodes $T^{[1]}_{s_1}$, $T^{[2]}_{n_0}$, $T^{[4]}_{n_1}$, $T^{[6]}_{n_2}$, and $T^{[7]}_{n_3}$.
Here, we assign the nodes with the dangling leg corresponding to the system indices and bath indices as the system nodes (red-shaded) and bath nodes (green-shaded), respectively. A TTNS can also have connecting nodes, which have exclusively connected legs, i.e., virtual indices, like nodes $T^{[3]}_X$ and  $T^{[5]}_X$. Therefore, the number of nodes $K$ can be larger than $N$. Every node can have none or multiple children nodes. For example, the first node $T^{[1]}$ has two children nodes, $T^{[2]}$ and  $T^{[3]}$.  $T^{[2]}$ has no children node.  Leaves are the nodes without children nodes.  The first system node is denoted as the root node, as shown in \Fig{TTNS_spinboson} (d). The root node has no parent node.  Other than the root node, every node has a parent node, which is the adjacent node on the path pointing toward the root node. For each node, all the legs are enumerated in a counter-clockwise manner with the bond connected to the parent node designated as the first leg, and the dangling leg (if it has) as the last leg, as shown exemplarily for node $T^{[5]}$ at the top-right corner of \Fig{TTNS_spinboson} (d). 

Ideally, strongly entangled tensors should be arranged as closely as possible. The tree shape in \Fig{TTNS_spinboson} (b) is in alignment with the entanglement topology of the super Hamiltonian $\mathcal{H}$ in \Eq{SuperHamiltonian} where all effective bath modes are directly coupled to the system DoF and they are independent of each other. However, the tensor network contraction cost for this tree structure scales as $\mathcal{O}(D_{\text{max}}^{P+1})$, which can be very expensive or even intractable when many effective environmental modes are taken into account. This issue can be resolved by inserting connecting nodes (yellow-shaded) between the system node and bath nodes, as illustrated in \Fig{TTNS_spinboson} (c) and (d). In the cases where every environmental mode imposes the same effect on the system dynamics, a balanced tree structure as in \Fig{TTNS_spinboson} (c) can be chosen, such that the distances between each bath node and the system node are equal. Otherwise, a more unbalanced tree structure can be adopted when the influence of some effective virtual phonon modes on the system dynamics is much stronger than the others. It is likely, that the impact of the effective phonon modes with a large Pad\'e pole number $p$ on the system dynamics is smaller than those with a smaller $p$. Thereby, in this work, we adopt the unbalanced TTNS in \Fig{TTNS_spinboson} (d) in our simulations.  

Recently, by taking advantage of the special structure of the HEOM method, Yan and Shi\cite{Yan_J.Chem.Phys._2021_p194104} proposed an efficient split operator method for propagating the binary TTNS. The method is implemented by successively constructing and updating the local MPS chain on the path connecting the system node and a bath tensor. In this work, we present an alternative route that directly employs the TDVP time propagation scheme on the whole TTNS.

\begin{figure*}
    \begin{minipage}[c]{0.25\textwidth} 		
	\raggedright a) \\
	\includegraphics[width=\textwidth]{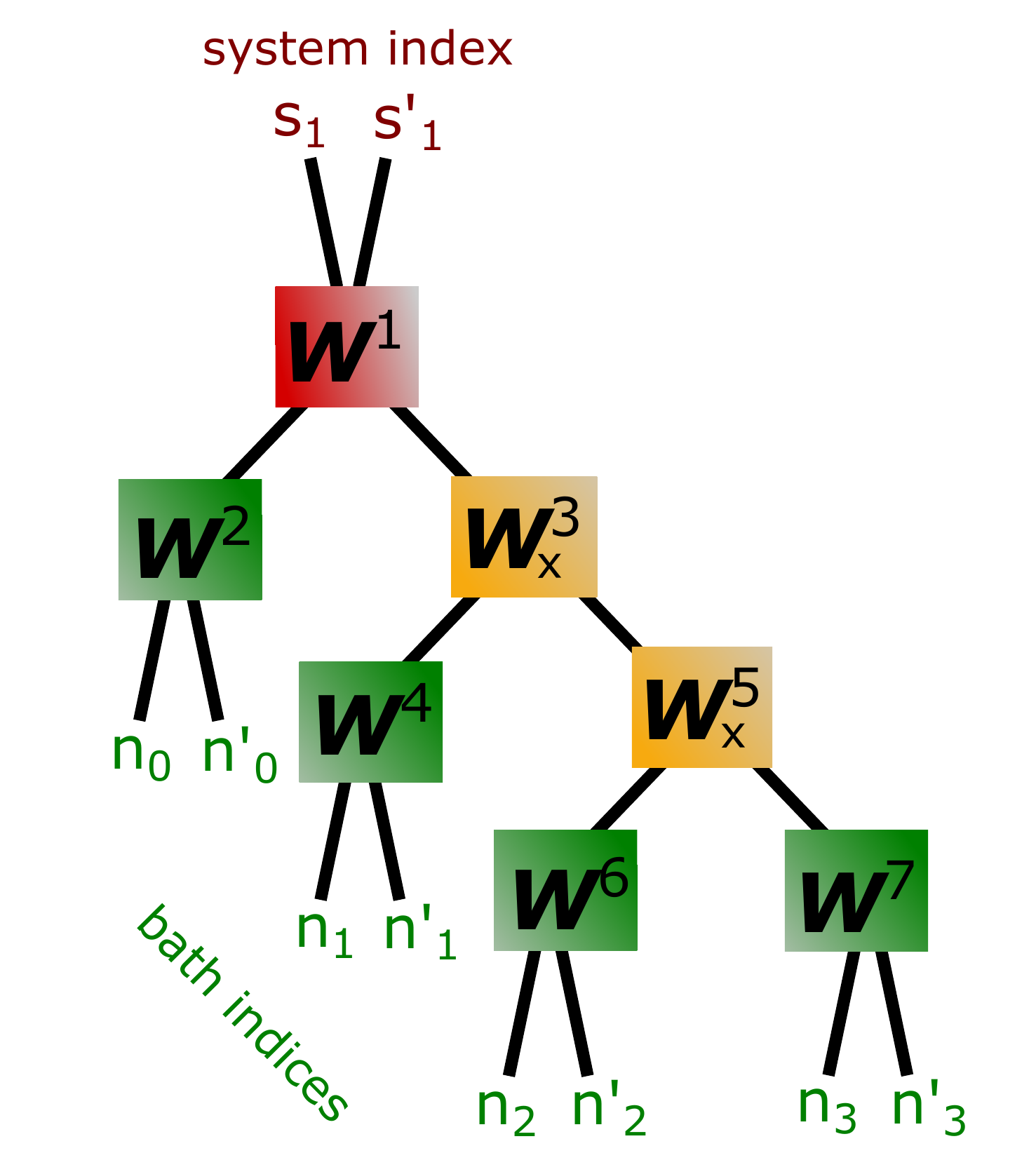}
    \end{minipage}
        \begin{minipage}[c]{0.2\textwidth} 		
	\raggedright b) \\
	\includegraphics[width=\textwidth]{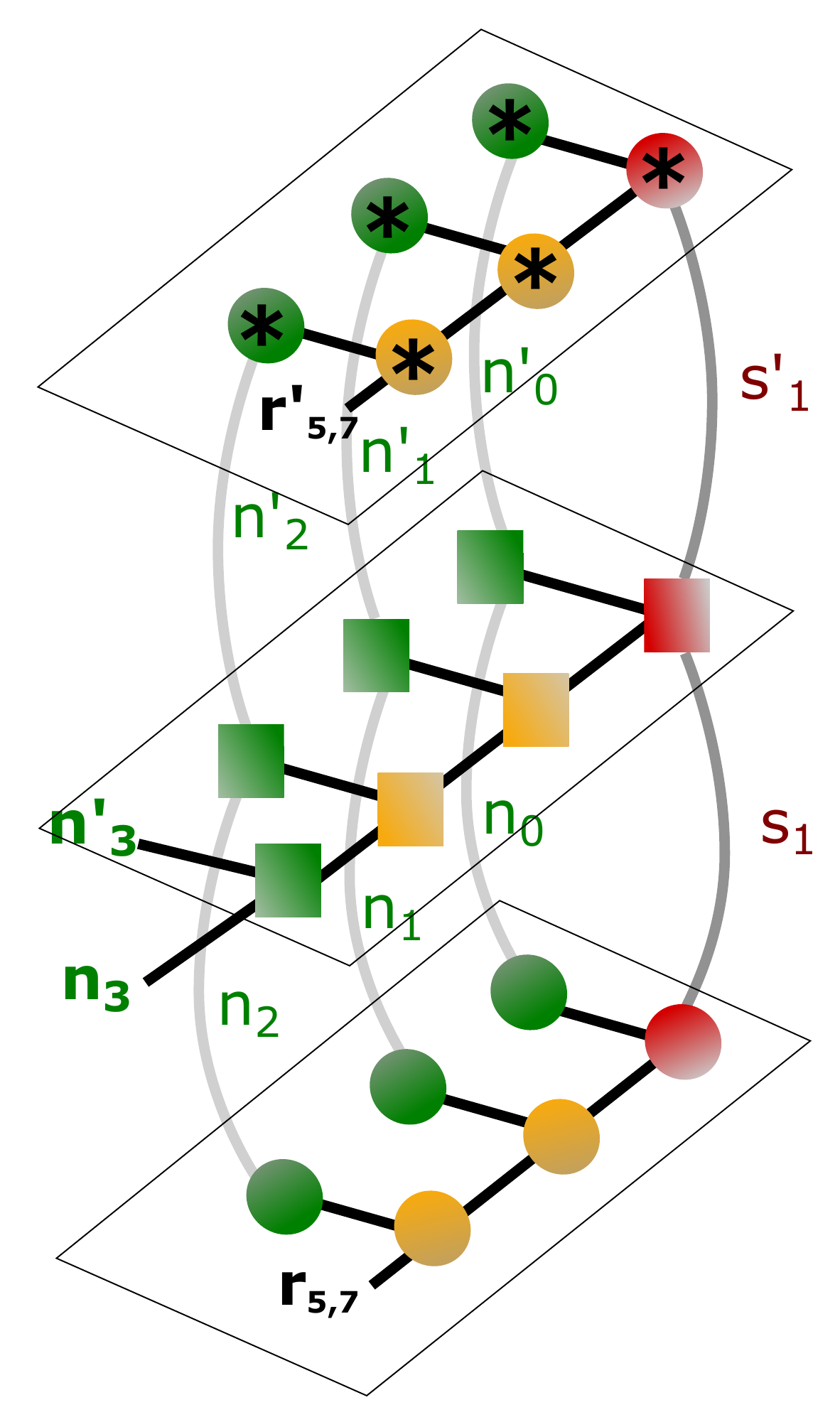}
    \end{minipage}
        \begin{minipage}[c]{0.2\textwidth} 		
	\raggedright c) \\
	\includegraphics[width=\textwidth]{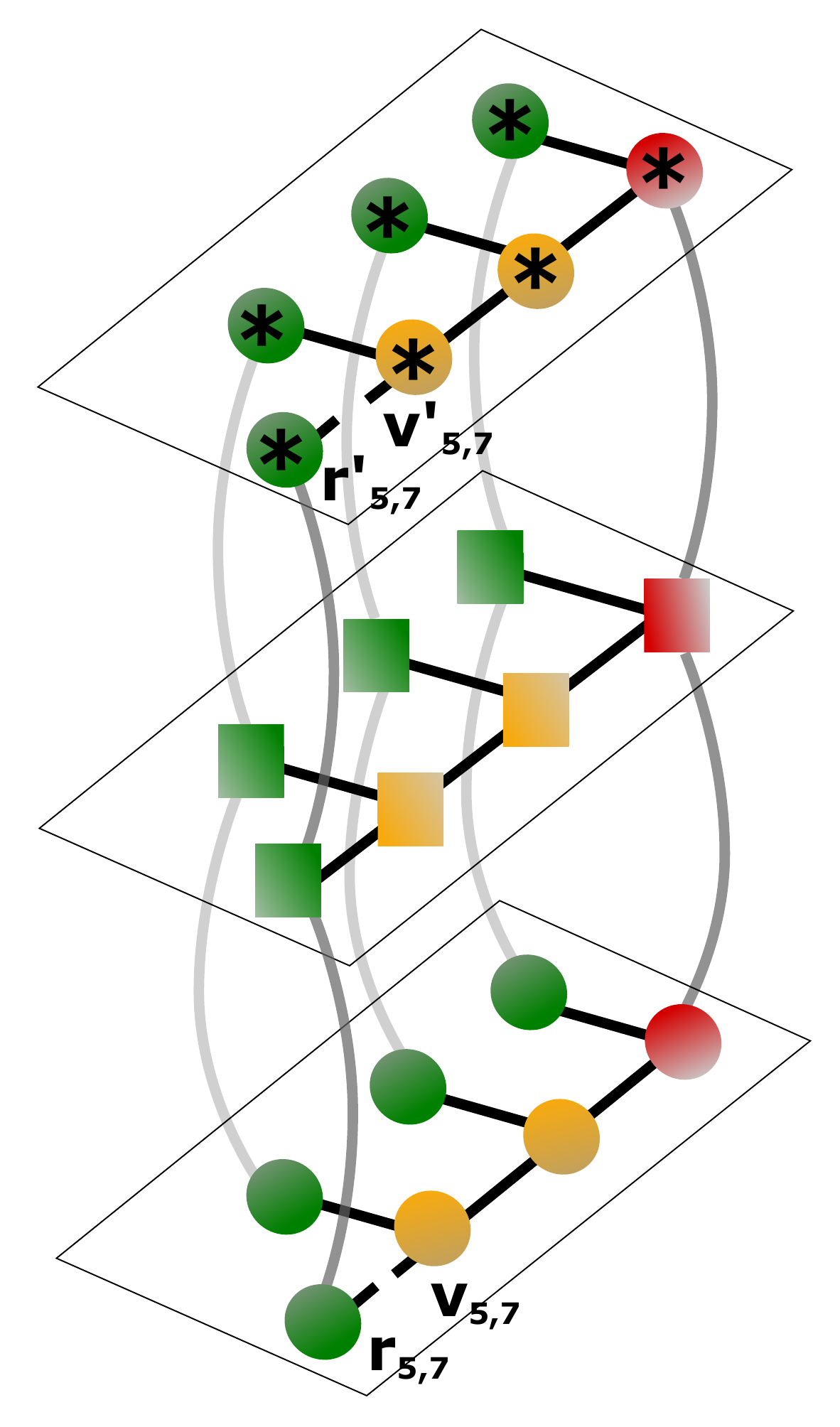}
    \end{minipage}
            \begin{minipage}[c]{0.2\textwidth} 		
	\raggedright d) \\
	\includegraphics[width=\textwidth]{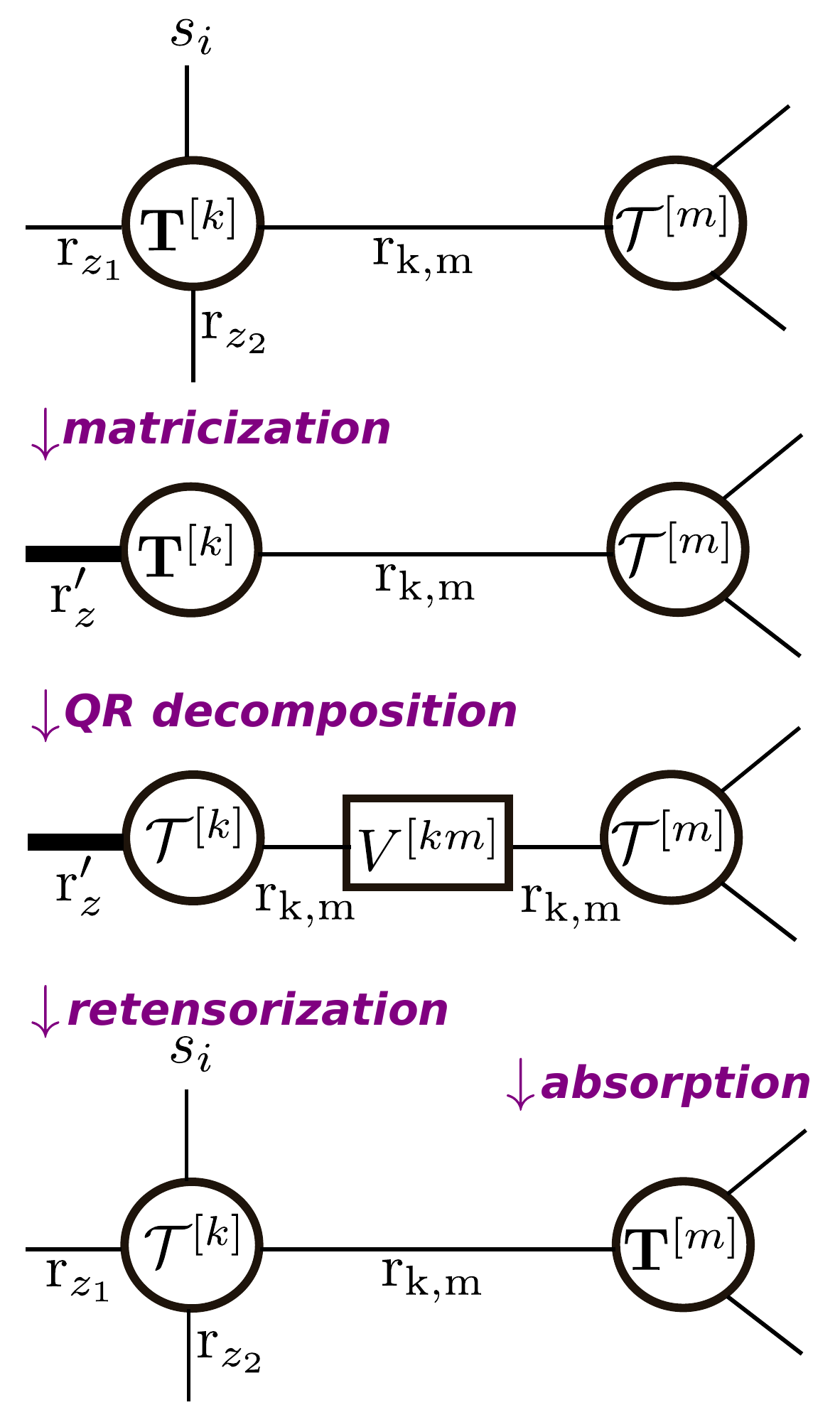}
    \end{minipage}
 \caption{(a) An example of the TTNO decomposition of the super Hamiltonian $\mathcal{H}$ (see \Eq{SuperHamiltonian}), which has the same network structure as the TTNS in \Fig{TTNS_spinboson} (b). To distinguish the TTNO from the TTNS in the graphical notation, each tensor in the TTNO is represented as a rectangle node. (b) and (c) display an example of the effective Hamiltonian $H_{[k]}^{\mathrm{eff}}$ (see \Eq{EffectiveHamiltonian_H}) for node $T^{[7]}$ and $G_{[km]}^{\mathrm{eff}}$ (see \Eq{EffectiveHamiltonian_G}) for the link between nodes  $T^{[5]}$ and $T^{[7]}$, respectively. The nodes with an asterisk denote taking their complex conjugation. (d) Schematic illustration of the tensor operations.}
 \label{TTNO_spinboson}	
\end{figure*}

To this end, the super Hamiltonian $\mathcal{H}$ in \Eq{SuperHamiltonian} needs to be decomposed into the product of tree tensor network operators (TTNO), which have the same network structure as the TTNS,
\begin{equation}
\label{superHamiltonian_TTNO}
\mathcal{H}= \mathrm{Contr}\left\{ W^{[1]}_{s_1,s'_1} W^{[2]}_{n_0,n'_0} \left(\prod_{p=1}^{P-1} W_X^{[2p+1]} W^{[2p+2]}_{n_p,n_p'}  \right) W^{[2P+1]}_{n_P,n'_P}\right\},
\end{equation}
as exhibited graphically in \Fig{TTNO_spinboson} (a) for $P=3$. The TTNO tensors $ W^{[1]}_{s_1,s'_1}$ and $ W^{[2p+2]}_{n_p,n'_p}$ in \Eq{superHamiltonian_TTNO} read 
\begin{equation}
\label{W1}
   W^{[1]}_{s_1,s'_1} =\left(
\begin{array}{cccc}
   \mathcal{H}_{S} & \hat{\sigma}_{z} & -\tilde{\sigma}_{z} &  \mathbbm{1}_4 \\
   \hat{\sigma}_{z}       &  & & \\
-\tilde{\sigma}_{z}       &  & & \\
      \mathbbm{1}_4            &  & &  \\
\end{array}
\right),
\end{equation} 
\begin{equation}
    W^{[2p+2]}_{n_p,n'_p}  =\left(
\begin{array}{c}
   \mathbbm{1}_L   \\
   \sqrt{\lambda} \left(b_{p}+ \eta_p b^{\dagger}_{p}  \right)      \\
\sqrt{\lambda} \left(b_{p}+ \eta^*_p b^{\dagger}_{p}  \right)       \\
      - i\gamma_{p} b^{\dagger}_{p} b_{p}      \\
\end{array}
\right),
\end{equation}
where the terms equal to zero have been left blank in \Eq{W1}. $\mathbbm{1}_L$ is an $L\times L$ unit matrix. $W_X$ is a rank-3 connecting tensor of size $4\times 4\times 4$ with the entries
\begin{equation}
   W_X[i,j,k]   =  \frac{\delta_{ij}\delta_{k=1}+\delta_{ik}\delta_{j=1}}{2\delta_{i=1}+\delta_{i\neq 1}}. 
\end{equation} 

With the TTNS and TTNO expression for the extended wave function $|\Psi(t)\rangle$ and the super Hamiltonian $\mathcal{H}$ explicitly given in Eqs. (\ref{TTNS}) and (\ref{superHamiltonian_TTNO}), respectively, we can use the TDVP-based time propagation algorithm to solve \Eq{SchroedingerEquation}. TDVP algorithm has been proven to be a stable and efficient time propagation scheme for the MPS/TT.\cite{Haegeman_Phys.Rev.B_2013_p75133,Lubich_SIAMJ.Numer.Anal._2015_p917,Haegeman_Phys.Rev.B_2016_p165116,Schroeder_Phys.Rev.B_2016_p75105,Baiardi_J.Chem.TheoryComput._2019_p3481,Mendl_arXivpreprintarXiv1812.11876_2018_p,Paeckel_Ann.Phys.NY_2019_p167998a}
The feasibility of the TDVP algorithm on the TTNS, the detailed derivation and related numerical analyses have been reported in Refs. \onlinecite{Bauernfeind_SciPostPhysics_2020_p024,Kloss_SciPostPhysics_2020_p70,Ceruti_SIAMJ.Numer.Anal._2021_p289}. Here, we only concisely discuss the implementation of the algorithm for our model and provide a pseudocode. 

In the one-site version of the TDVP algorithm, the Schr\"odinger equation in \Eq{SchroedingerEquation} is solved by projecting the wave function into the so-called tangent space, i.e., a manifold of all TTNSs with fixed bond dimensions, \cite{Bauernfeind_SciPostPhysics_2020_p024,Kloss_SciPostPhysics_2020_p70,Ceruti_SIAMJ.Numer.Anal._2021_p289}
\begin{equation}
\label{TDVP_SLE}
i\frac{|\Psi[T]\rangle}{\partial t} = \mathcal{P}
_{\mathcal{M}\{\Psi[T]\}} \mathcal{H} |\Psi[T]\rangle.
\end{equation}
The explicit expression of the tangent space projection operator $\mathcal{P}
_{\mathcal{M}\{\Psi[T]\}}$ can be found in Refs. \onlinecite{Bauernfeind_SciPostPhysics_2020_p024,Kloss_SciPostPhysics_2020_p70,Ceruti_SIAMJ.Numer.Anal._2021_p289}. By using Trotter breakups, the projected time-dependent Schr\"odinger equation in \Eq{TDVP_SLE} can be integrated locally, where each tensor $T^{[k]}$ and the link $V^{[km]}$ between node $k$ and one of its neighboring node $m$ in a specified sequence are updated according to the following equations 
\begin{equation}
\label{local_node}
\frac{\partial T^{[k]}(t)}{\partial t} = -i H_{[k]}^{\mathrm{eff}} T^{[k]}(t), 
\end{equation}
\begin{equation}
\label{local_link}
\frac{\partial V^{[km]}(t)}{\partial t} = i G_{[km]}^{\mathrm{eff}} V^{[km]}(t), 
\end{equation}
while all other tensors are fixed. The effective local Hamiltonian $H_{[k]}^{\mathrm{eff}}$ in \Eq{local_node} and $G_{[km]}^{\mathrm{eff}}$ in \Eq{local_link} are given by
\begin{equation}
\label{EffectiveHamiltonian_H}
\begin{split}
H_{[k]}^{\mathrm{eff}} = \mathrm{Contr}'_k&\left\{T^{[1]}_{s_1} W^{[1]}_{s_1,s'_1} T^{[1]*}_{s'_1} \cdots  T^{[k-1]} W^{[k-1]}T^{[k-1]*} W^{[k]} \right.\\
&\left. T^{[k+1]} W^{[k+1]}T^{[k+1]*} \cdots T^{[K]}_{n_P} W^{[K]}_{n_P,n'_P}T^{[K]*}_{n'_P}\right\},
\end{split}
\end{equation}
\begin{equation}
\label{EffectiveHamiltonian_G}
\begin{split}
G_{[km]}^{\mathrm{eff}} = \mathrm{Contr}{''}_{km}&\left\{T^{[1]}_{s_1} W^{[1]}_{s_1,s'_1} T^{[1]*}_{s'_1} \cdots  T^{[k]} W^{[k]}T^{[k]*} \right.\\
&\left.\cdots T^{[K]}_{n_P} W^{[K]}_{n_P,n'_P}T^{[K]*}_{n'_P}\right\}.
\end{split}
\end{equation}
Here, we use the Einstein convention that the duplicate indices are summed over. The asterisk denotes taking the complex conjugation of the tensor. $\mathrm{Contr}'_k$ denotes contracting over all physical and virtual indices in the TTNS decomposition of $C_{\bm{s}}^{\bm{n}}(t)$ and $C_{\bm{s}}^{\bm{n}*}(t)$  as well as the TTNO expression for $\mathcal{H}$, except for leaving the indices connected to $T^{[k]}$ and $T^{[k]*}$ and physical indices to $W^{[k]}$ open. For $\mathrm{Contr}{''}_{[km]}$, the bond connecting $T^{[k]}$ and $T^{[m]}$ as well as the bond connecting $T^{[k]*}$ and $T^{[m]*}$ are left open. As an example, we show the graphical representation of $H_{[7]}^{\mathrm{eff}}$ and $G_{[57]}^{\mathrm{eff}}$ in \Fig{TTNO_spinboson} (b) and (c), respectively.

A full TDVP step starts from a canonicalized TTNS, where all tensors except for the root node are orthonormalized, $\bm{T}^{[1]}(t), \mathcal{T}^{[2]}(t), \cdots, \mathcal{T}^{[K]}(t)$, and is accomplished when all tensors are propagated to $\bm{T}^{[1]}(t+\Delta t), \mathcal{T}^{[2]}(t+\Delta t), \cdots, \mathcal{T}^{[K]}(t+\Delta t)$ after a sweeping along the entire network, as illustrated by the orange dotted lines in \Fig{TTNS_spinboson}. A full sweeping procedure starts and ends at the root node, and every node is entered and exited $n_c^k+1$ times where $n_c^k$ is the number of the children nodes to node $k$.  For example, a sweep over the TTNS in \Fig{TTNS_spinboson} (d) goes through the nodes sequentially, and the path is $1\rightarrow 2\rightarrow 1 \rightarrow 3 \rightarrow 4 \rightarrow 3 \rightarrow 5 \rightarrow 6 \rightarrow 5 \rightarrow 7 \rightarrow 5 \rightarrow 3 \rightarrow 1$. 
To keep track of the time evolution, we introduce a variable $h_k$ and it is initialized as $h_k=0$ before the sweeping procedure starts. Moving from node $k$ to a neighboring node $m$ along the direction of the sweeping path, the following steps are implemented:
\begin{enumerate}
    \item Evolve $\bm{T}^{[k]}(t+h_k)$ forward in time according to
    \begin{equation}   \bm{T}^{[k]}\left(t+h_k+\frac{\Delta t}{n_c^k+1}\right) = e^{-i H^{\mathrm{eff}}_{[k]}\frac{\Delta t}{n_c^k+1}}\bm{T}^{[k]}(t+h_k),
    \end{equation}
    and then update $h_k = h_k+\frac{\Delta t}{n_c^k+1}$. $H^{\mathrm{eff}}_{[k]}$ is computed with $\mathcal{T}^{[1]}(t+h_1), \cdots, \mathcal{T}^{[k-1]}(t+h_{k-1}), \mathcal{T}^{[k+1]}(t+h_{k+1}), \cdots, \mathcal{T}^{[K]}(t+h_K)$.
    \item Permute the indices of $\bm{T}^{[k]}(t+h_k)$ and then reshape the tensor into a matrix such that the index $r_{k,m}$ corresponding to the bond connecting nodes $k$ and $m$ now serves as the column of the matrix, while all other indices are grouped into one index as the row of the matrix; Conduct the QR decomposition of the matricized $\bm{T}^{[k]}(t+h_k)$  into $\mathcal{T}^{[k]}(t+h_k)$ and $V^{[km]}(t+h_k)$; Retensorize $\mathcal{T}^{[k]}(t+h_k)$ as a new local tensor on node $k$.
    \item Evolve $V^{[km]}(t+h_k)$ in time according to
    \begin{equation}
    V^{[km]}\left(t+h_m\right) = e^{i G^{\mathrm{eff}}_{[km]}(h_k-h_m) }\,V^{[km]}(t+h_k),
    \end{equation}
    where $G^{\mathrm{eff}}_{[km]}$ is computed with $\mathcal{T}^{[1]}(t+h_1), \cdots, \mathcal{T}^{[k]}(t+h_k),  \cdots, \mathcal{T}^{[K]}(t+h_K)$.
    \item Absorb $V^{[km]}\left(t+h_m\right)$ into $\mathcal{T}^{[m]}(t+h_m)$ to obtain  $\bm{T}^{[m]}(t+h_m)$. 
\end{enumerate}
The tensor operations in the second and fourth steps are schematically illustrated in \Fig{TTNO_spinboson} (d). After implementing the above steps, the orthogonality center in the TTNS is also moved from node $k$ to the adjacent node $m$. 

\begin{figure}
  \begin{minipage}[c]{0.35\textwidth} 		
  	\raggedright a) 
	\includegraphics[width=\textwidth]{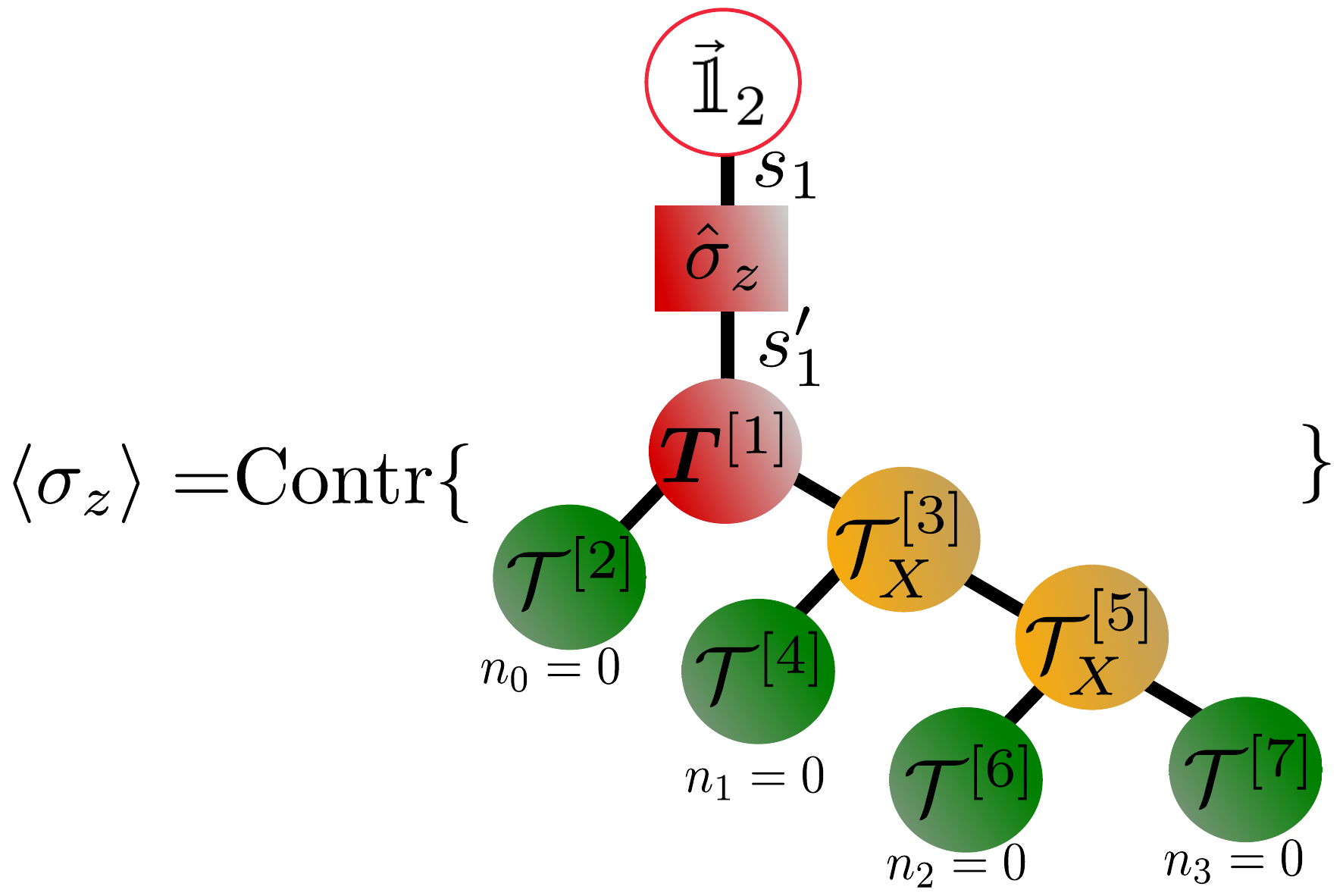}
    \end{minipage}
    \begin{minipage}[c]{0.45\textwidth} 		
    	\raggedright b) \\
	\includegraphics[width=\textwidth]{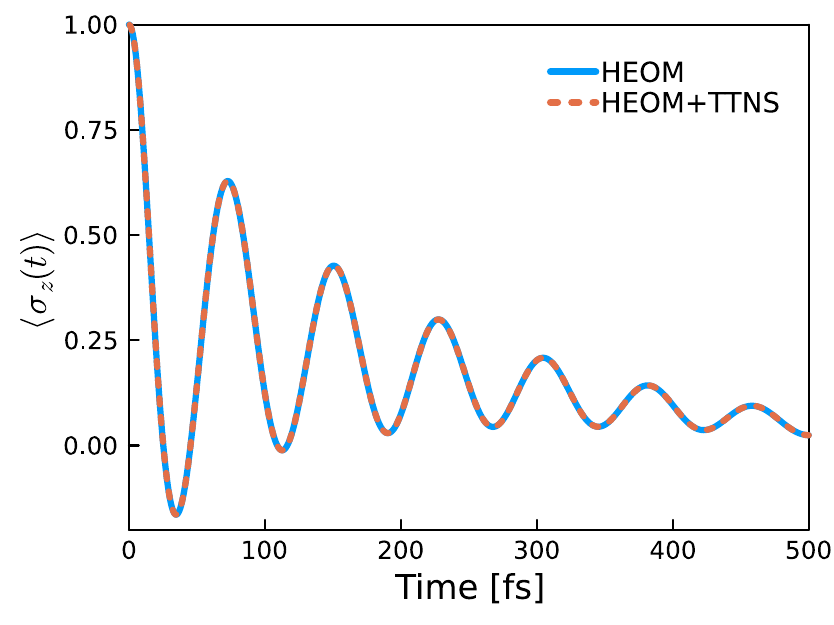}
    \end{minipage}
 \caption{(a) Graphical illustration of calculating $\langle \sigma_z(t)\rangle$ (see \Eq{sigmaz_TTNS}) with the TTNS shown in \Fig{TTNS_spinboson} (d). (b) Dynamics of the spin-boson model calculated using the conventional HEOM with a hierarchical truncation tier $L=50$ and the HEOM+TTNS method with a maximal bond dimension $D_{max}=5$. The following parameters are used: $\epsilon=50 \text{ cm}^{-1}$, $\Delta=200 \text{ cm}^{-1}$, $\beta^{-1}=300K$, $\lambda=100\text{ cm}^{-1}$, and $\Omega=10\text{ cm}^{-1}$.}
 \label{results_spinboson}	
\end{figure}

To assess the accuracy and performance of the proposed HEOM+TTNS method against the conventional HEOM method, we first conduct a benchmark calculation on the spin-boson model. 
The spin is initially placed in the spin-up state. The parameters can be found in the caption of \Fig{results_spinboson}. Note that, we adopt a much smaller value for $\Omega$ than other parameters, which indicates that the bath relaxation is slow as compared to other system dynamical processes, and the non-Markovian feature in this case can be strong.  
We inspected the spin dynamics $\langle \sigma_z(t)\rangle=\mathrm{tr}\{\sigma_z \rho(t)\}$ and found that the converged result using the conventional HEOM method is obtained with a very deep truncation tier $L=\sum_{p=0}^P n_p=50$ and $P=3$. 
Using the HEOM+TTNS method with the TTNS displayed in \Fig{TTNS_spinboson} (d), $\langle \sigma_z(t)\rangle$ is calculated by
\begin{equation}
\label{sigmaz_TTNS}
\begin{split}
\langle \sigma_z(t)\rangle 
=\mathrm{Contr} &\left\{\vec{\mathbbm{1}}_{2}[s_1] \hat{\sigma}_z[s_1,s'_1] \bm{T}^{[1]}_{s'_1}   \right. \\
& \mathcal{T}^{[2]}_{n_0=0}\left(\prod_{p=1}^{P-1}  \mathcal{T}^{[2p+1]}_X\mathcal{T}^{[2p+2]}_{n_p=0} \right) \left. \mathcal{T}^{[2P+1]}_{n_P=0}\right\},
\end{split}
\end{equation}
which is graphically illustrated in \Fig{results_spinboson} (a). For every effective bath mode, we take $0\le n_p\le L$, which means that actually more ADOs are taken into account than in the conventional HEOM method. We found that with a very low maximal bond dimension $D_{max}=5$, the result obtained through the HEOM+TTNS method is already in excellent agreement with that obtained through the conventional HEOM method, as shown in \Fig{results_spinboson} (b).

For a fair comparison, all the codes are written using Julia programming language and our simulations are run on a single CPU core of an Apple MacBook Pro laptop with M1 chip. For the conventional HEOM method, 
we use an exponential integrator as proposed in Ref. \onlinecite{Wilkins_J.Chem.TheoryComput._2015_p3411}, which requires less memory usage as compared to Runge-Kutta methods.
 Every simulation step takes on average 6.7 seconds with a time step $\Delta t= 0.025\text{ fs}$ and requires a memory of 7.6 GB.  Using the HEOM+TTNS method with the previously stated TDVP integration scheme, we gain a significant speed-up in computational time and save in memory usage. Every simulation step takes only 0.028 seconds in CPU time and memory usage of 10 MB. Besides, it is also found that the maximally allowed time step in the HEOM+TTNS method is $\Delta t=0.5\text{ fs}$, much larger than that used in the conventional HEOM method. We should mention that our code for the conventional HEOM method is not fully optimized. The introduction of the filtering algorithm proposed by Shi \etal\cite{Shi_J.Chem.Phys._2009_p84105} would greatly improve its efficiency. However, the simulation time using the HEOM+TTNS method grows nearly linearly with the increase of pole number $P$ and truncation tier $L$, in contrast to the factorial scaling in the conventional HEOM method. As such, even for the relatively simple spin-boson model, we found that the simulation using the HEOM+TTNS method is generally much faster than the conventional HEOM method.
 
 We also tested the MPS/TT decomposition of the extended wave function $|\Psi(t)\rangle$ for the above model (see the tensor network structure in \Fig{TTNS_spinboson} (a)) and the matrix product operator format of the super Hamiltonian $\mathcal{H}$ is obtained using ITensor Package.\cite{ITensor}  The same time propagation algorithm and code as in the HEOM+TTNS method are used.
 For this model, although we found that the result of the HEOM+MPS/TT method (data not shown) also converges with a maximal bond dimension of $D_{max}=5$, every simulation step is four times slower than that of the HEOM+TTNS method. This implies the pure influence of a tree shape in determining the practical computational cost.

As a proof-of-concept, we demonstrated in the main text the algorithm and results specific to the spin-boson model. The method can be generalized to much more complicated systems. In the SI, we show the applicability of the HEOM+TTNS method for the Frenkel exciton model which is widely used in studying the excitation energy transfer in photosynthetic systems,\cite{Ishizaki_ProceedingsoftheNationalAcademyofSciences_2009_p17255} as well as a spin-boson-like chain model where each spin is coupled to its own bosonic environment and the nearest-neighbor spins. 

In summary, we presented the feasibility of adopting a genuine TTNS and TDVP-based time propagation algorithm for solving the HEOM method, and compared its performance to the conventional HEOM method as well as the HEOM-MPS/TT scheme.  Because the TTNS inherits many properties from the MPS/TT formalism, we expect that the proposed HEOM+TTNS method can also benefit from algorithmic optimizations and advances that have been developed in the context of the MPS/TT representation.\cite{Yang_Phys.Rev.B_2020_p94315,Secular_Phys.Rev.B_2020_p235123,Dunnett_Phys.Rev.B_2021_p214302,Borrelli_J.Phys.Chem.B_2021_p5397,Li_arXivpreprintarXiv2208.10972_2022_p,Xu_JACSAu_2022_p335,Ceruti_SIAMJ.Numer.Anal._2023_p194} Besides, we also hope that the idea presented in this work may inspire more applications of the tensor network states in combination with the HEOM method to enable a more efficient simulation of a much broader class of chemical and physical systems. For example, we will extend the current approach to study open fermionic system problems in our future work.\cite{ke2023current,Evers_Adv.Mater._2022_p2106629}

\section*{Acknowledgements}
The author thanks M. Thoss for helpful discussions. This work was supported by the German Research Foundation (DFG).

\section*{Supplementary Material}
See the supplementary material for the implementation details  of the HEOM+TTNS method for the Frenkel exciton model and a spin-boson-like chain model.

\section*{Data Availability}
The data and code that support the findings of this study are available from the corresponding author upon reasonable request.

%
 \end{document}